\documentclass[a4paper,12pt]{article}  
\usepackage{array}
\usepackage{enumitem}
\usepackage{graphics}
\usepackage{eurosym}
\usepackage[noabbrev]{cleveref}

\newcolumntype{L}[1]{>{\raggedright\let\newline\\ \arraybackslash\hspace{0pt}}m{#1}}
\usepackage[left=2cm,right=2cm,top=2cm,bottom=2cm]{geometry}
\usepackage{graphicx}
\let\OLDthebibliography\thebibliography
\renewcommand\thebibliography[1]{
  \OLDthebibliography{#1}
  \setlength{\parskip}{0pt}
  \setlength{\itemsep}{0pt plus 0.3ex}
}
\usepackage{xspace} 

\def\urltilda{\kern -.15em\lower .7ex\hbox{\~{}}\kern .04em}
\def\urldot{\kern -.10em.\kern -.10em}
\def\urlhttp{http\kern -.10em\lower -.1ex\hbox{:}\kern -.12em\lower 0ex\hbox{/}\kern -.18em\lower 0ex\hbox{/}}

\def\gm2{{\tt g-2}\xspace}

\def\Mu3e{{\tt Mu3e}\xspace}

\def\comet{{\tt COMET}\xspace}

\def\muec{{$\mu^{-}{N} \rightarrow e^{-}{N}$}}

\newcommand{\gbp[1]}{\pounds\kern 0.08333em{#1}\xspace}
\newcommand{\eu[1]}{\euro\kern 0.08333em{#1}\xspace}
\newcommand{\chf[1]}{SFr.\kern 0.08333em{#1}\xspace}
\newcommand{\usd[1]}{\$\kern 0.08333em{#1}\xspace}
\newcommand{\mueg}{$\mu^{+} \rightarrow e^{+} \gamma$\xspace}
\newcommand{\meee}{$\mu^{+} \rightarrow e^{+}e^{+}e^{-}$\xspace}
\newcommand{\metoee}{$\mu^{-}+e^{-} \rightarrow e^{-} + e^{-}$\xspace}

\newcommand{\mue}{$\mu \to e$\xspace}

\newcommand{\mupc}{$\mu^{-}$$-$$e^{+}$\xspace}

\begin{document}
%
\mbox{}
\vskip 5cm
\pagestyle{empty}

\begin{center} \begin{sffamily} \begin{bfseries}
{\Huge COMET}
\end{bfseries} \end{sffamily} \end{center}

\begin{center}
{\large 
J.-C. Ang\'elique, C.~C\^arloganu, W.~da~Silva, A.~Drutskoy, M.~Finger, D.~N.~Grigoriev, T.~Kachelhoffer, F.~Kapusta, Y.~Kuno\footnote{contact person: kuno@phys.sci.osala-u.ac.jp.}, P.~Lebrun, R.~P.~Litchfield, D.~Lomidze, D.~Shoukavy, A.~M.~Teixeira, I.~Tevzadze, Z.~B.~Tsamalaidze, 
Y.~Uchida, V.~Vrba, K.~Zuber}
\end{center}

\begin{center}
  {\large 
A submission to the 2020 update of the European Strategy for Particle Physics
on behalf of the COMET collaboration. }
\end{center}

\begin{center} \begin{sffamily} \begin{bfseries}
{\Large Abstract}
\end{bfseries} \end{sffamily} \end{center}

The search for charged lepton flavour violation (CLFV) has enormous discovery potential in probing new physics Beyond the Standard Model (BSM). The observation of a CLFV transition would be an undeniable sign of the presence of BSM physics which goes beyond non-zero masses for neutrinos. Furthermore, CLFV measurements can provide a way to distinguish between different BSM models, which may not be possible through other means.
So far muonic CLFV processes have the best experimental sensitivity because of the huge number of muons which can be produced at several facilities world-wide, and in the near future, new muon beam-lines will be built,  
leading to increases in beam intensity by several orders of magnitude.
Among the muonic CLFV processes, \mue conversion is one of the most important processes, having several advantages compared to other such processes.

We describe the COMET experiment, which is searching for \mue conversion in a muonic atom at the J-PARC proton accelerator laboratory in Japan. The COMET experiment has taken a staged approach; the first stage, COMET Phase-I, is currently under construction at J-PARC, and is aiming at a factor 100 improvement over the current limit. The second stage, COMET Phase-II is seeking another 100 improvement (a total of 10,000), allowing a single event sensitivity (SES) of $2.6 \times 10^{-17}$ with $2\times 10^{7}$ seconds of data-taking. Further improvements by one order of magnitude, which arise from refinements to the experimental design and operation, are being considered whilst staying within the originally-assumed beam power and beam time.
Such a sensitivity could be translated into probing many new physics constructions up to $\mathcal{O}(10^{4})$\,TeV energy scales,
which would go far beyond the level that can be reached directly by collider experiments. 
The search for CLFV \mue conversion is thus highly complementary to BSM searches at the LHC.



\newpage

\pagestyle{plain}
\pagenumbering{arabic}
\setcounter{page}{1}

\bigskip
\begin{sffamily} \begin{bfseries}
  {\Large Scientific Context}
\end{bfseries} \end{sffamily} 


\bigskip

The observation of neutrino oscillations provided the first laboratory
evidence of a phenomenon which could not be accounted for by the
Standard Model of Particle Physics (SM). In particular, 
neutrino oscillations imply that neutrinos are massive, and that
individual lepton flavours are not conserved. This contradicts 
the original SM formulation, in which neutrinos are massless by
construction, and an (accidental) symmetry leads to the conservation
of total and individual lepton numbers. 
This departure from the SM paradigm---together with other
observations suggesting the need for New
Physics (the lack of a viable dark matter
candidate and the inability to account for the observed baryon
asymmetry of the Universe)---further implies that numerous other
processes, forbidden in the SM, might indeed occur in Nature. 

The violation of flavour conservation in the neutral lepton sector
opens the door to the interesting possibility of CLFV. Being strictly forbidden in the SM, once the
latter is minimally extended by Dirac right-handed neutrinos 
to account for neutrino oscillation data, CLFV rare transitions and
decays can occur at loop level, mediated 
by light massive neutrinos
and $W^\pm$ bosons; nevertheless, the expected rates are extremely
small, lying beyond any conceivable experimental sensitivity.
For example, the rate of the CLFV radiative muon decay is found to be 
BR(\mueg)$\sim \mathcal{O}(10^{-54})$. 

The positive experimental observation of any CLFV process thus signals the
presence of a New Physics (NP) model, a true departure from the SM,
even minimally extended via massive Dirac neutrinos. 

In the presence of BSM physics, many CLFV rare transitions and decays
can occur, in addition to the above mentioned one. There
is a vast array of observables, including purely leptonic processes (as for
example radiative decays $\ell_i \to \ell_j \gamma$ and 
three-body decays $\ell_i \to 3\ell_j$), transitions occurring
in muonic atoms (such as neutrinoless muon-electron conversion), 
CLFV leptonic and
semi-leptonic meson decays, as well as flavour-violating $Z$ and Higgs
boson decays, among others. Many appealing 
BSM constructions, 
from minimal extensions to UV complete models, do predict
contributions to CLFV  observables which are either already ruled out
by current bounds, or that lay within the expected sensitivity of
near-future facilities. Especially in view of the so-far negative
results of direct searches for BSM being carried at the LHC, indirect
searches for rare transitions (at the ``high-intensity'' frontier) 
have very strong potential.
 
Firstly, the observation of a CLFV process may constitute
the first (albeit indirect) discovery of New Physics; moreover, and
as substantiated by extensive phenomenological analyses, 
CLFV searches can probe BSM regimes---masses and
couplings---well beyond the
reach of direct searches at the high-energy frontier. 
Secondly, the contributions to the different CLFV processes 
strongly reflect the nature of the New Physics model at work, i.e.
the new interactions and properties of the new mediators:
while radiative decays are always loop processes, three-body
decays, or \mue 
conversion in the presence of nuclei, can, remarkably, occur at the tree
level. A full picture of the contributions to different CLFV
observables can also be instrumental in shedding light on the nature of
the BSM interaction: photonic or nonphotonic, and in the latter 
case also on the type of current---(pseudo)scalar, (axial)vector or
even tensor. 

Clearly, CLFV observables have huge potential in 
disentangling 
new physics in the leptonic sector. Although it is important to
stress that CLFV need not arise from the mechanism responsible for
neutrino oscillations (and hence for flavour mixing in the neutral
lepton sector), this remains a very appealing possibility; many
well-motivated mechanisms of neutrino mass generation---in particular
realisations at comparatively light scales---are expected to lead to
sizeable contributions to the CLFV observables, as well as to the
correlation of certain decay modes. Should there be a connection
between neutrino masses and low-energy lepton flavour violation, then
the CLFV observables will be crucial in putting it to the test, and
hopefully unveiling the mechanism at work. 

Currently, the muon system is one of the
best laboratories to look for CLFV, and hence for BSM models 
capable of (observable) contributions to the above mentioned rare
decays and transitions~\cite{muonCLFV}.   
Muons can be abundantly produced (providing the high
statistics required to
investigate processes having very small rates); intense 
muon beams can be obtained at meson factories when low energy protons 
$E_p < 1$~GeV hit light targets (as at PSI, TRIUMF, LANL), as well 
as  at proton accelerators (such as J-PARC or Fermilab), where muons
are created as by-products of high-energy collisions. Moreover, 
the ``relatively long'' lifetime
of the muon makes it possible to manipulate them into optimal experimental configurations. 
Its low mass further implies that the number of kinematically
allowed decay channels, flavour-conserving or not, is relatively
small; the simple final states can also be measured with great
precision. 

Radiative CLFV muon decays, $\mu^+ \to e^+ \gamma$, have been searched for
since the 1940's. The current bound on these decays is 
BR(\mueg)$< 4.2 \times 10^{-13}$, obtained by the MEG
Collaboration at PSI~\cite{TheMEG:2016wtm}. In the future, MEG II is expected to
improve the sensitivity to $ 6 \times 10^{-14}$~\cite{Baldini:2018nnn}
(see also~\cite{Cavoto:2017kub}). 
The three-body muon decay, \meee also offers
excellent prospects to look for CLFV. At present, the best bound is
still that of SINDRUM II~\cite{Bellgardt:1987du},  
BR(\meee)$< 1.0 \times 10^{-12}$,
expected to be significantly improved in the coming years by the Mu3e
collaboration at PSI to around $10^{-15}$~\cite{Blondel:2013ia}, possibly
$10^{-16}$, should very high-intensity muon beams become available~\cite{muonCLFV}.

Many interesting CLFV processes can be studied when muons are trapped
and form so-called ``muonic atoms''. When negatively-charged muon
beams hit a target, a muon can be stopped, and then cascaded down in energy
until it effectively forms a $1s$ bound state. Normally, it then decays
in orbit ($\mu^- \to e^- \nu_\mu \bar \nu_e$), 
or is captured by the nucleus,  $\mu^- + (A,Z) \to \nu_\mu+(A,Z-1)$.
In the presence of NP,
one of the most interesting CLFV processes which can occur 
is neutrinoless muon-to-electron conversion,
\begin{equation}
\mu^- \,+ \,(A,Z) \,\to \,e^-\, + \,(A,Z)\,,
\end{equation}
in which $(A,Z)$ denotes the mass and atomic numbers of the target
nuclei. 
The event signature of coherent \mue conversion in a muonic atom
is the emission of a mono-energetic single electron with
an energy ($E_{\mu e}$) of 
$E_{\mu e} = m_{\mu}-B_{\mu}-E_\mathrm{recoil}$, 
where $m_{\mu}$ is the muon mass, $B_{\mu}$ is the binding energy of
the $1s$-state muonic atom, and $E_\mathrm{recoil}$ 
denotes the nuclear recoil energy,
which is small. Since $B_{\mu}$ varies for various nuclei, $E_{\mu e}$
will also be different depending on the material in which the muon stops. The
nuclear recoil energy is approximately given by $E_{\rm recoil} = (m_{\mu}
- B_{\mu})^2/(2m_{N})$, where $m_{N}$ is the mass of the recoiling
nucleus, and is typically small: for instance, $E_{\mu e}$ = 104.97\,MeV for
Aluminium (Al), $E_{\mu e}$ = 104.3\,MeV for Titanium (Ti) and $E_{\mu
e} = 94.9$\,MeV for Lead (Pb).
It is also important to emphasise that 
after the conversion, the target nucleus can be left either in
the ground state, or in one of its excited states. 
In general, the transition to the
ground state (called coherent capture) is dominant, since the rate of
the latter over the non-coherent one is enhanced by
a factor approximately equal to the number of nucleons in the nucleus
(since all of the nucleons participate in the process).
Also the time distribution of the occurrence of \mue conversion depends on 
the lifetime of muonic atoms for a particular nucleus (0.864 microseconds for Aluminium).

From an experimental point of view, neutrinoless 
\mue conversion is one of the most attractive
CLFV processes. Firstly, the $e^{-}$ energy of about 105~MeV is well above the
end-point energy of the free muon decay spectrum ($\sim
52.8$~MeV). Secondly, since the event signature is a mono-energetic
electron, no coincidence measurement is required. Thus the search for
this process has the potential to improve sensitivity by using a high
muon rate without suffering from accidental background events, which is
a serious problem for \mueg decay searches and other muon CLFV
processes, such as \meee decays.


The current experimental limit is CR$(\mu^{-}+{\rm Au} \to e^{-} + {\rm Au}) < 7 \times 10^{-13}$, obtained by SINDRUM-II at PSI~\cite{SINDRUM_MUEN}.
Here, CR is the rate of the \muec~conversion
process normalised to the normal muon nuclear capture process.
In the future, several experiments will be
dedicated to looking for muon-electron conversion: 
DeeMe~\cite{Nguyen:2015vkk} aims at reaching a sensitivity of $10^{-14}$ for SiC targets; working with
Aluminium targets, Mu2e at Fermilab~\cite{Bartoszek:2014mya} expects to reach 
$< 7 \times 10^{-17}$  at 90\,\% CL and its upgraded experiment, Mu2e-II~\cite{Mu2e-II-EOI}, aims at a factor of ten or more than Mu2e. 
At J-PARC, the goal of the COMET experiment
is to reach $<7 \times 10^{-15}$ at 90\,\% CL in its Phase-I~\cite{comet-phaseI}, 
and ultimately, a final
Phase-II sensitivity of $\mathcal{O}$$(10^{-17})$~or better~\cite{loi, proposal, cdr, Krikler:2016gdq, technote164}, as described in detail later.

Clearly, searches for \mue conversion will allow probing the
presence of CLFV sources with unprecedented sensitivity. 
New Physics 
contributions can be generically divided 
into dipole-photonic contributions and nonphotonic (or
contact) interactions; in the former case, the mediator is a photon
which is absorbed by the nucleus, while in the latter 
the CLFV interaction is due to
the exchange of new heavy virtual particles that couple both to the
leptons and to the quarks in the nucleus.  As mentioned before,
a wide variety of New Physics models can give rise to 
short-distance (nonphotonic) CLFV interactions. Unlike radiative \mueg
muon decays, which are only sensitive to electromagnetic dipole
interactions, the \mue conversion process could be
mediated by (pseudo)scalar, (axial)vector, or tensor
currents.
Recent studies~\cite{Crivellin:2017rmk} support that
while the current experimental bounds on \mueg are
clearly powerful in constraining the dipole operators, 
neutrinoless \mue conversion is the most
sensitive observable 
to explore operators involving quarks; it also appears to be
the best setup to study (most) vector interactions. Future
experimental prospects might even render it more sensitive than \mueg to the
dipole operators. 
At present, 
the three-body decay, \meee remains the most powerful observable to
explore and constrain four-fermion operators with $\mu eee$ flavour
structure
(unless the sensitivity to \mue conversion improves to
$\mathcal{O}(10^{-18})$), which could well be the case.

While in most studies only coherent \mue conversion processes 
have been considered, CLFV tensor and axial-vector four-fermion
operators could also contribute to the process. Since the latter
operators couple to the spin of nucleons, they can therefore mediate
a spin-dependent conversion. 
Since---as noticed before---the contributions from dipole, scalar and
vector operators are spin-independent, the rate difference between the
spin-dependent and spin-independent \mue conversion processes 
could be used to discriminate the tensor and axial-vector operators
from the others. Notice however, that while spin-independent 
\mue conversion is a coherent process whose rate is enhanced as 
$A^2$ ($A$ being the number of nucleons), 
the spin-dependent conversion does not have the coherent enhancement.
This further suggests that a high experimental sensitivity is
required, as achievable by the COMET experience.
The Aluminium target, which will be used by the COMET Phase-I
experiment, has a nuclear spin of 5/2. 
Therefore, if a \mue conversion signal is observed, other targets
with zero nuclear spin should be studied to investigate which BSM
operators are at work. 

An impressive number of phenomenological studies have highlighted the
probing power of neutrinoless \mue conversion. 
Numerous UV complete extensions of the SM (such
as supersymmetry, large extra dimensions, extended Higgs sectors, ...) can easily account for
large 
contributions to \mue conversion, within the sensitivity of COMET 
(some associated with very distinctive correlations with other CLFV 
observables). Likewise, SM extensions
including right-handed (sterile) neutrinos, as in the case of
low-scale realisations of the type I seesaw and its variants, also lead to
significant contributions. In all these cases, it has been argued that  
searches for \mue conversion might provide a unique window into
regimes which might not be directly accessible at colliders. 
While in the above mentioned SM extensions \mue conversion is
typically a loop process (via photon, $Z$ or $H$ penguins, or box
diagrams), many BSM models can induce tree level contributions, from
scalar and vector field exchanges. This is the case of type III
seesaw realisations, extended Higgs sectors, and R-parity violating SUSY
models. 
Interestingly, some of the most appealing and successful explanations
to the recently observed anomalies in B-meson decays call upon NP
states which are expected to have a strong impact on CLFV. This is
the case of BSM constructions 
including additional neutral vector bosons
($Z^\prime$), and of (scalar) leptoquark models
(fields carrying both hadron and lepton numbers). In both cases,
contributions to muon-electron conversion arise at the tree level, and in
numerous realisations this observable is associated with the most (if not
the most!) stringent constraints on the scale of the mediators and the
size of their (CLFV) couplings, allowing to 
probe regimes beyond direct collider
reach.

The physics case for experiments dedicated to look for \mue conversion
is far from being limited to searches for neutrinoless \mue conversion. These 
experiments can be 
adapted to search for the lepton number violating (LNV) mode,
\begin{equation}
\mu^{-} + N(A,Z) \rightarrow e^{+} + N^{\prime}(A,Z-2)\,, 
\end{equation}
if one can (directly) determine 
whether the emitted lepton is an electron or a positron; this is the
case of COMET Phase-I (by virtue of its lack of charge selection in
the final state), and of Mu2e.
The sensitivity to the LNV mode should be in the same ballpark, 
typically $\mathcal{O}(10^{-14}-10^{-16})$. 
The current experimental bound was obtained by SINDRUM II (for
Titanium atoms), and is
$\mathrm{CR}(\mu^{-} + \mathrm{Ti} \to e^{+} + \mathrm{Ca}) \,<\, 
1.7 \times 10^{-12}~ (3.6 \times 10^{-11})
$~\cite{Dohmen:1993mp},
where 
the numbers denote the limit obtained at 90\%
C.L. for a transition to the ground state (to the excited states through giant dipole resonance) of 
Calcium.
Notice that in the case of an LNV conversion 
the final state nucleus (different from the initial one) can
be either in the ground or in an excited state. Having different
initial and final state nuclei further precludes a coherent
enhancement of the transition amplitude---which implies that it will
not be augmented in large atoms. The experimental signal is also less
clean than that of the coherent conversion: not only the emitted
positron is no longer monoenergetic, but there are new important sources of
background, such as pions and protons. 

The ability to search (and discover) this very rare LNV transition is
extremely relevant in view of the important connection it might
establish to the presence of BSM Majorana mediators that are
responsible for the violation of total lepton number. These new
states are an integral part of many models of neutrino mass
generation, and might also play 
a relevant role in explaining the baryon asymmetry of the Universe via
leptogenesis, which requires---among other ingredients---LNV processes.

Another observable that can strengthen the physics programme of COMET
is the CLFV Coulomb enhanced decay of the muonic atom into a pair of
electrons, 
$\mu^-\,e^- \to e^-\,e^-\,$~\cite{Koike:2010xr}.
From an experimental point of
view, it presents several advantages  with respect to other muon
channels, such as \mueg, or \meee.  
When compared to the three-body muon decay, the
new observable has a larger phase space, and a cleaner experimental
signature, consisting of back-to-back electrons with a well defined
energy ($\sim m_\mu/2$).  The rate receives a 
strong enhancement due to the Coulomb interaction with the nucleus,
which scales as $(Z-1)^3$, thus leading to larger branching ratios for Lead or
Uranium atoms.
The $Z$ dependence of the $\mu^{-}e^-\to e^-e^-$ 
amplitude, as well as the angular and energy 
distribution of the emitted electrons, are sensitive to the
underlying nature of the BSM interaction (photonic vs. contact), thus
being useful in distinguishing New Physics models~\cite{Uesaka:2017yin}.

\bigskip
While CLFV muon decays and conversion may offer the most impressive
experimental sensitivity, they do not offer a complete view into the
full picture of CLFV. In order to do so, one must aim at exploring as
many CLFV observables as possible---not only in the muon sector, but
also in tau decays (as well as in Higgs and $Z$-boson CLFV decays). 
Due to its comparatively large mass, tau leptons decay 
via numerous leptonic and semi-leptonic modes, and may include several that violate charged lepton flavour.
In particular, CLFV tau decays allow direct access to $\tau\to e$ and
$\tau\to\mu$ flavour violation. 
Due to its larger mass (and thus larger Yukawa couplings), 
tau decays could be particularly sensitive to BSM contributions
proportional to the decaying lepton mass, as for example 
Higgs-mediated CLFV transitions.

However, when compared to the muon sector, CLFV $\tau$ 
searches are hampered by the comparatively shorter lifetime 
($\tau_\tau = 2.91 \times 10^{-13}$\,s), 
which prevents the realisation of tau-beams. 
The required high-luminosity (or large production) 
can be nevertheless  
obtained via $\tau^+\tau^-$ pair production in $e^+e^-$
storage rings. 
Several of these facilities are typically optimised for studying
B-meson physics, and the best current bounds on tau CLFV decays have been
in fact obtained at BaBar and
Belle~\cite{Amhis:2016xyh,Lusiani:2018zvr}, and are
$\mathcal{O}(10^{-7} -10^{-8})$. 
In the future, Belle II is expected to improve the bounds 
by 1-2 orders of magnitude. Likewise, and due to the large amounts of
tau leptons produced at the LHC, LHCb, ATLAS and CMS are also expected
to improve the existing bounds.


\vspace{5mm}

%

\begin{sffamily} \begin{bfseries}
{\Large Methodology}
\end{bfseries} \end{sffamily}

\bigskip

The COMET experiment (J-PARC E21) is seeking to observe coherent neutrinoless \mue conversion in a muonic atom, \muec, at the Japan Proton Accelerator Research Complex (J-PARC) in Tokai, Japan~\cite{loi,proposal,cdr}. 
COMET stands for COherent Muon to Electron Transition. 
The single-event sensitivity (SES) reach expected for COMET Phase-II is  $2.6 \times 10^{-17}$ for $2 \times 10^{7}$ seconds of data-taking~\cite{loi,proposal,cdr}. It is a factor of 10,000 better than the current limit of $7\times 10^{-13}$ which was obtained by SINDRUM-II at PSI~\cite{SINDRUM_MUEN}.  It is noted that the COMET Phase-II muon beam line would provide about $2 \times 10^{11}$ stopped muons per second, which would be the highest muon beam intensity in the world.  


\begin{figure}[htbp!]
  \begin{center}
   \includegraphics[width=0.7\textwidth]{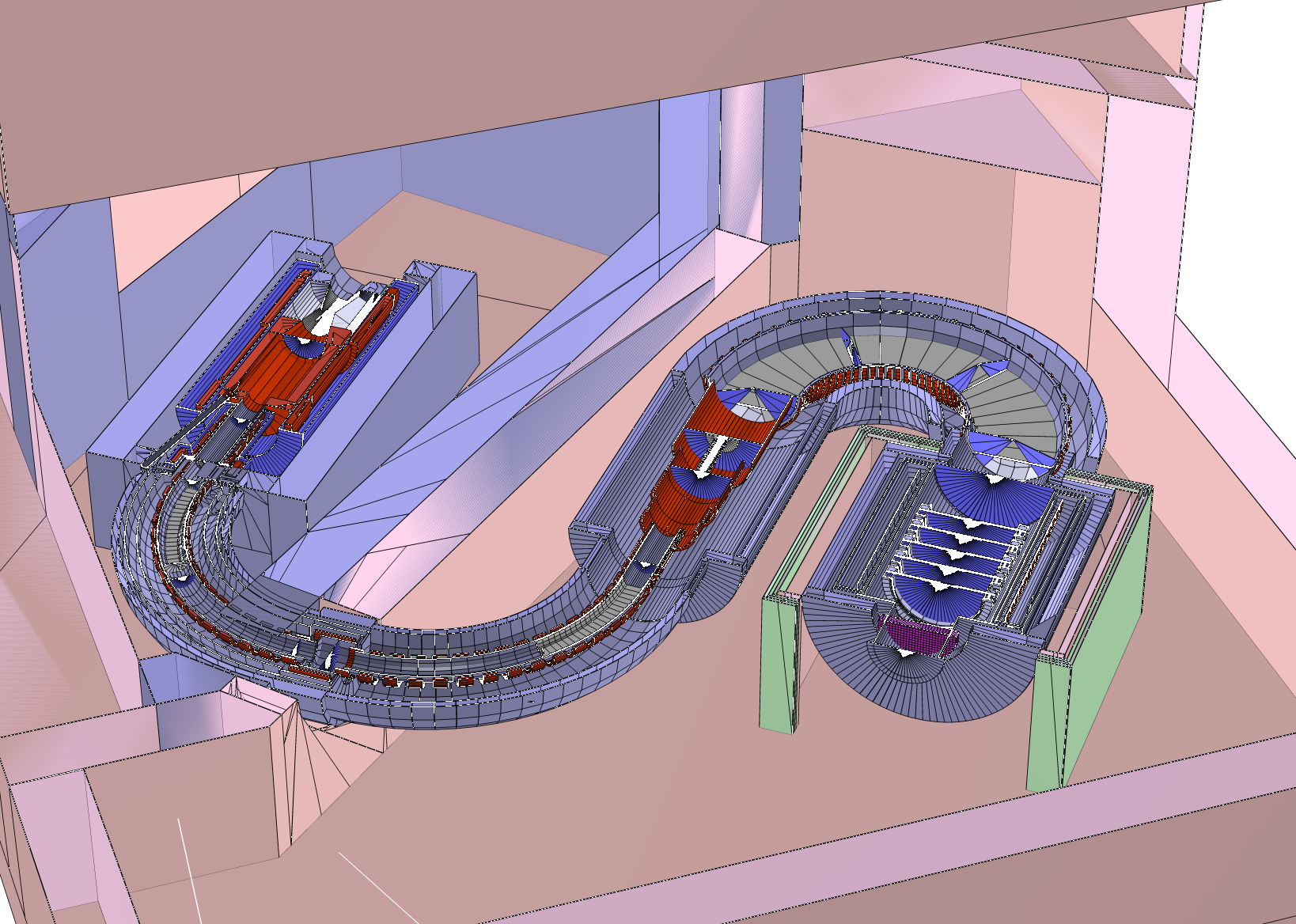}
  \end{center}
  \caption{\it{\small{A cutaway view of the full Phase-II layout of the COMET experiment, showing the pion capture solenoid (on the left), the muon transport beam line 
  with tunable momentum selection as the muons travel towards the muon stopping target, and the electron spectrometer (on the right), 
also tunable, which removes neutral and wrong-sign particles as well as selecting the momentum of the particles which travel through to the detector section in the foreground.}}   \label{figure:cometphaseii} }
\end{figure}

The COMET experiment will make use of a dedicated 8\,GeV, 7\,$\mu$A proton beam (with a power of 56 kW), which is slow-extracted from the J-PARC Main Ring (MR), via a new proton beamline to the J-PARC Nuclear and Particle Physics Experimental hall (NP Hall). A schematic layout of the COMET setup is shown in \Cref{figure:cometphaseii}. 

Muons will be produced from the pions generated in the collisions of the 8\,GeV protons with a production target made of tungsten.  
The yield of low momentum muons transported to the experimental area is enhanced using a superconducting 5\,T pion capture solenoid surrounding the proton target in the pion capture section in \Cref{figure:cometphaseii}. 
Muons are momentum- and charge-selected using 180$^{\circ}$ curved superconducting solenoids in the muon transport section of 3\,T, before being stopped in an Aluminium target located in the target section.  
The signal electrons from the muon stopping target are transported through the electron spectrometer composed by the 180$^{\circ}$ curved superconducting solenoids and are detected by instrumentation in the detector section in a 1\,T magnetic field. The curved electron spectrometer will be used to eliminate low-energy background electrons and transport the signal electrons to the detector section with high efficiency.

There are several potential sources of electron background events in the
energy region around 100~MeV, which can be grouped into three
categories as follows: the first group includes intrinsic physics backgrounds
which come from muons stopped in the target; the second 
corresponds to
beam-related backgrounds which are caused by beam particles (muons) and
other 
particles possibly contaminating the muon beam (pions, electrons, anti-protons and so on); the third includes 
cosmic-ray induced backgrounds, fake tracking events and so
on. 
Among the first group, the most serious one is electrons from bound muon decays in orbit (DIO) of a muonic atom, where the energy spectrum rises beyond $m_{\mu}/2$ owing to nuclear recoil, and the endpoint of the spectrum coincides with the energy of $E_{\mu e}$.  These DIO background events can be eliminated only by measurements with high momentum resolution.

To suppress the occurrence of beam-related background events, a pulsed proton beam, where proton leakage between pulses is kept extremely low, is adopted. Since a muon in a muonic atom of Aluminium has a lifetime of $0.864\,\mu$s, a pulsed beam with a shorter beam duration compared to this lifetime, and a beam repetition comparable to or longer than the lifetime, would allow the removal of prompt beam background events, by using a delayed time window to make the measurements.  Such a bunched proton beam can be realised at the J-PARC MR by bunched slow-extraction. This has been experimentally demonstrated, resulting in a beam with pulse spacing of 1.17 microseconds.
There are stringent requirements on the proton beam leakage during the measurement interval. We have also demonstrated that the number of leakage protons with respect to the number of protons in the beam pulse (proton extinction) is of ${\cal{O}}(10^{-11})$ in the MR, which meets the COMET requirement. We are further testing proton extinction at the beam lines in the experimental hall.
The low proton energy available at the J-PARC MR also allows for excellent beam extinction between pulses. 
%

The proton beam energy of about 8~GeV has been chosen in order to reduce anti-proton production which might cause some background events. The proton energy can be lowered further for COMET, if further reduction of anti-protons is required. This is one advantage of the dedicated use of the J-PARC MR.


To eliminate cosmic-ray induced background events, both passive and active shielding will be used. The passive shielding consists of concrete, polyethylene and lead.  Active shielding is provided by a scintillator-based veto system together with a resistive plate chamber system, covering the whole muon beamline and detector sections.

The detector for signal electron detection is a combination of a straw-tube tracker and an electron calorimeter (ECAL) with fast-scintillating LYSO crystals, and is called the StrECAL. Instead of a conventional multiple over-woven layer method, the straws in COMET are made of a single layer, rolled and attached itself in a straight line using ultrasonic welding. 
The straws for COMET Phase-II are 5\,mm in diameter and 12\,$\mu$m in wall thickness.  A momentum resolution of better than 200\,keV/$c$ is expected for the straw-tube tracker. The ECAL covers the cross-section of the 50-cm radius detector region and 1920 crystals are needed. Each crystal has a $2\times2$ cm$^2$ cross-section and 12\,cm in length (10.5 radiation length). They are read by avalanche photodiodes with an active area of $10\times10$\,mm$^2$.  The expected energy resolution of less than 5\,\% at about 100\,MeV has been demonstrated in a prototype test. This provides the momentum and energy resolutions required to discriminate between signal and backgrounds at a sensitivity of $10^{-18}$.


COMET is based on fundamental principles that originated in the MELC experimental proposal\footnote{Mu2e is more similar to the MELC design.}~\cite{MELC},  but with some significant differences, as described in the following: the curved solenoid sections for COMET are equipped with dipole coils which superimpose a vertical magnetic field on them; this allows the momentum of the particles which travel preferentially through the centre of the solenoids to be varied. 
The feasibility of this method of muon production and momentum selection has been proven at the MuSIC facility at the Research Center for Nuclear Physics (RCNP) in Osaka University~\cite{MUSIC}. 

The ability to tune the momentum of the muons impinging on the muon-stopping target by altering the field in the first 180$^{\circ}$ curved solenoid in the muon beam transport will help to better understand and to further reduce the systematic effects and backgrounds affecting the measurement.
The second 180$^{\circ}$ curved solenoid, which makes up the Electron Spectrometer section, ensures that there is no line of sight between the target and the detector systems. It eliminates all neutral particles from the muon stopping target hitting the detectors. 
The curved Electron Spectrometer also collimates away (in a tunable way) the numerous charged particles which are produced at momenta outside the signal region, such as muon-decay electrons.

The above described design differences with respect to MELC$-$in particular the tunable dipole fields and the curved electron spectrometer section which are unique to COMET$-$allow Phase-II to have a high potential sensitivity to \muec, and to achieve this ultimate sensitivity in a timely manner.

\vspace{-5mm}
\paragraph{Phased Deployment}

\begin{figure}[htbp!]
  \begin{center}
   \includegraphics[width=0.7\textwidth]{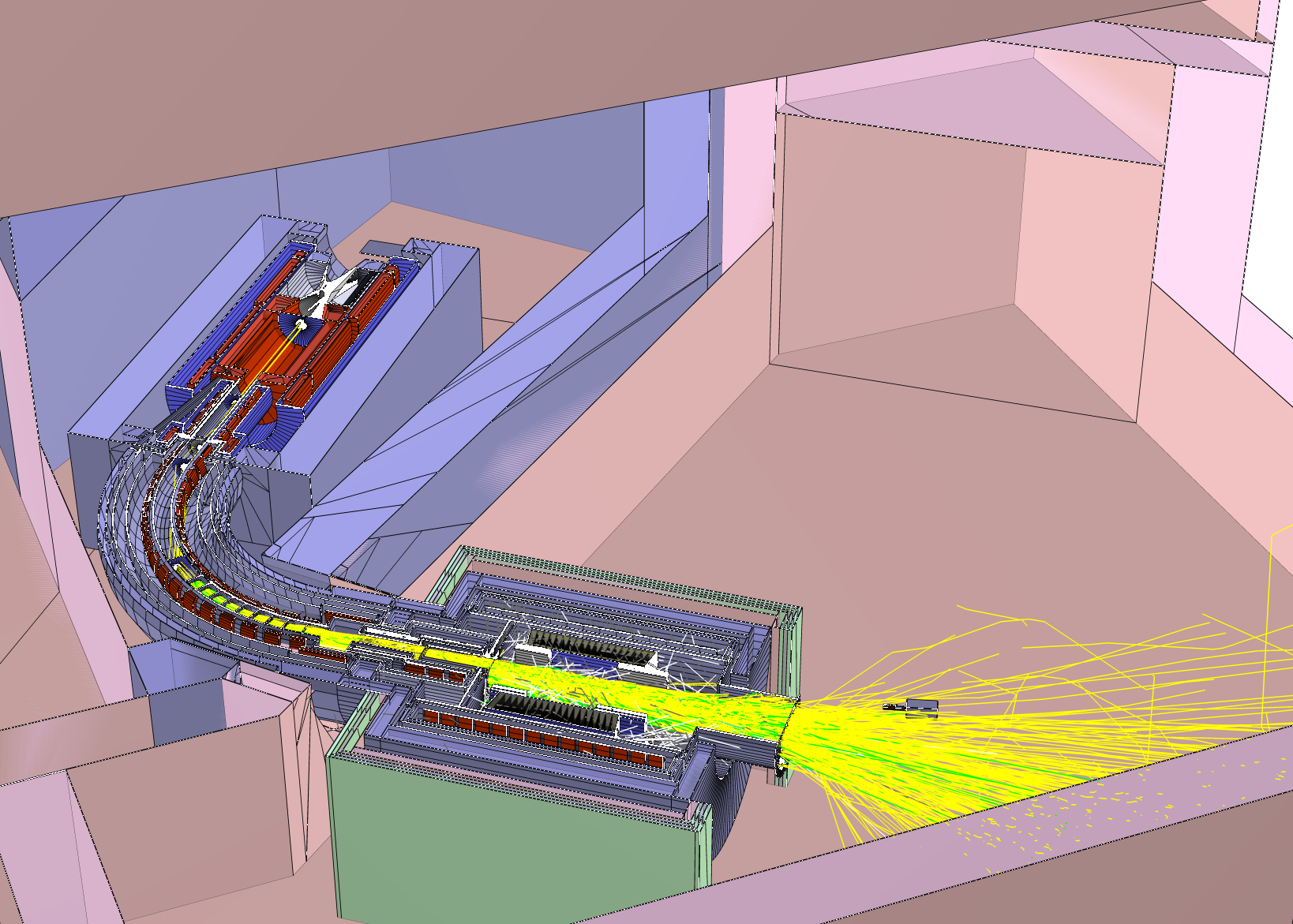}
  \end{center}
  \caption{\it{\small{A cutaway view of the Phase-I layout of the COMET experiment, showing the first 90$^{\circ}$ bend of the muon transport beam line. The detector will be placed at the end of the muon transport section.}}} 
  \label{figure:cometphasei} 
\end{figure}

The COMET Collaboration has opted to use a staged approach to experiment deployment, to ensure that detailed measurements of this new muon beam production facility can be made (in the form of COMET Phase-I~\cite{comet-phaseI}), before embarking on the full COMET configuration (COMET Phase-II).
The Phase-I facility will have the pion capture section and the muon transport section up to the end of the first 90$^{\circ}$ bend. The detectors will be installed after the end of this 90$^{\circ}$ bend. The layout of COMET Phase-I is shown in \Cref{figure:cometphasei}.
COMET Phase-I has the dual goal of studying the novel muon production beam line such that it is fully understood in preparation for Phase-II, and of making measurements of \mue conversion with a sensitivity that is approximately 100 times better than the previous limit, at a SES of 3$\times 10^{-15}$.
COMET Phase-I will utilise a 8-GeV proton beam of 0.4\,$\mu$A, yielding a beam power of 3.2 kW. The pion production target is made of graphite, instead of the tungsten used in Phase-II.
With a total number of protons on target (POT) of $3.2 \times 10^{19}$ (which corresponds to about 150 days), about $1.5 \times 10^{16}$ muons in total will be stopped, which is sufficient to reach the design single event sensitivity of COMET Phase-I. 

The primary COMET Phase-I detector for searching for the neutrinoless \mue
conversion signals is composed of a cylindrical drift chamber (CDC) and a set of trigger
hodoscope counters, which together are called the CyDet.
The CDC, with a total of about 5000 sense wires in stereo views has been constructed and is being tested at KEK.

The experimental setup of COMET Phase-I will be augmented with prototypes of the Phase-II straw-tube tracker and the electron calorimeter, called the StrECAL detector. The straw tubes used in Phase-I are 9.75\,mm in diameter and have 20\,$\mu$m thick walls.
The detector magnet is a solenoid of 1\,T, where either the CyDet or StrECAL detector is placed at any time.

As well as providing valuable experience with the detectors, the StrECAL and CyDet 
will be used to characterise the backgrounds to the signal of neutrinoless \mue conversion to ensure that the Phase-II SES can be realised.


\vspace{-5mm}
\paragraph{Phase-II : }
The initial SES of the COMET Phase-II was $2.6 \times 10^{-17}$ for $2\times 10^{7}$ seconds of data-taking with 56\,kW proton beam power from J-PARC MR~\cite{loi,proposal,cdr}. This represents a factor of 10,000 improvement over the current limit. This original design is very conservative in terms of the high proton beam power available at J-PARC.
Recently, the COMET collaboration has refined the experimental design and operation of the COMET Phase-II. It was shown that even with the same beam power and the beam time as originally assumed, the sensitivity can be potentially further improved by one order of magnitude, down to $\mathcal{O}(10^{-18})$. Possible improvements include the design of the electron spectrometer as well as the proton and muon targets~\cite{technote164}.


The Phase-II set-up requires the construction of the second half of the muon beam transport and the Electron Spectrometer, but is otherwise composed of a reconfiguration of many of the active parts of Phase-I. In particular, the high-radiation region near the pion production section, which is a cost-driver, will be built to the Phase-II specifications (for 56\,kW beam power) from the beginning. 
This will allow a smooth continuation from COMET Phase-I to COMET Phase-II.

\vspace{-5mm}
\paragraph{Timeline : }%
The J-PARC proton beam will arrive at the COMET experimental area in early 2020, when Phase-I beam studies and integration will commence, and  Phase-I physics data-taking and analysis will follow. 
By the mid-2020s, it is expect that the full Phase-II experiment will be deployed and running.
If \muec\, is observed, the COMET Phase-II will try to measure it with different target materials, for instance up to medium-heavy nuclei, to identify which effective interaction is responsible for it~\cite{Kitano:2002mt,Cirigliano:2009bz}.
The Phase-II muon beam facility, with providing $2\times 10^{11}$ muons per second, will also be the world's most intense pulsed muon source.
Moreover, with its double curved solenoid and dipole field configuration, the Phase-II muon beam facility will produce extremely high-quality beams of variable momentum.

The completed COMET Phase-II configuration can be adapted to search for and measure several CLFV and LNV processes, in addition to the main \muec\,channel. These include \mupc\,~\cite{Yeo:2017fej} and \metoee conversions~\cite{Koike:2010xr, Uesaka:2017yin, Uesaka:2016vfy} as well as \mueg, with the addition of photon converters. A broad programme of study is expected to continue well beyond 2025 and into the 2030s, with a specific path that is dependent on the observations that will have been made by that time.
Some of these additional measurements will require the beam line to run in dedicated positive-muon mode, which will produce an extremely high-quality beam in the Phase-II configuration.

\vspace{-5mm}
\paragraph{European Contributions : }
Currently, approximately 35 institutions participate to COMET Phase-I, from Australia, {\bfseries Belarus}, China, {\bfseries Czech Republic}, {\bfseries France}, {\bfseries Georgia}, {\bfseries Germany}, India, Japan, Kazakhstan, South Korea, Malaysia, {\bfseries Russia}, {\bfseries United Kingdom}, and Vietnam. There is further growing interest in each of the regions that are represented.

The European contributions to \comet include: \vspace{-3mm}
\begin{itemize}
\item{Cosmic Ray Veto detector (Belarus, France, Georgia, Russia)} \vspace{-3mm}
\item{Electromagnetic calorimeter (Belarus, Russia)} \vspace{-3mm}
\item{Muon target monitor (Germany)} \vspace{-3mm}
\item{Data-acquisition and detector triggering systems (UK, Czech Republic)} \vspace{-3mm}
\item{Straw-tube tracking detector (Georgia, Russia)} \vspace{-3mm}
\item{Muon stopping targetry (Germany)} \vspace{-3mm}
\end{itemize}

Controlling and monitoring the beam composition and the various backgrounds for this rare-decay experiment requires very large simulated data samples. Single- and multi-bunch simulations in Europe have involved significant contributions in terms of CPU (France, UK, and Germany), storage and data sharing (France) and production management (UK). Software developments related to the analysis, track finding and track fitting optimisation have lead also to intensive software tests and improvements (UK, France, Germany). In particular, much effort has been focused on introducing simulation strategies that allow for high-statistics background and signal estimates without requiring a proportional increase in computational resources. Combining such strategies with increasing international resource contributions will allow the computational challenges of COMET to be met.

\vspace{-5mm}
\paragraph{Further Physics Measurements : }
Some COMET collaboration members are also heavily involved in the R\&D towards the PRISM project. 
PRISM stands for Phase-Rotated Intense Slow Muon source.
It would provide a high flux, monochromatic muon beam with highly-suppressed pion backgrounds. They can be achieved by a muon storage FFAG ring with the novel technique of beam phase-rotation, where slow muons are accelerated and fast muons are decelerated by RF in the storage ring. A long flight path in the muon storage ring effectively eliminates any remaining pions in a beam down to order of $10^{-20}$. The prototype of the FFAG storage ring was constructed at Osaka University to test the principle of operation of muon phase rotation.
PRISM combines the advantages of the COMET Phase-II configuration with a muon storage ring, and when combined with the J-PARC proton beam power upgrade to 1.3\,MW and higher, it has a potential sensitivity of the order of $10^{-19}$. 

PRISM will provide an ideal way to make measurements on multiple choices of muon-stopping materials, including much heavier elements than Aluminium and possibly spin-dependent \mue conversion searches~\cite{Cirigliano:2017azj, Davidson:2017nrp}.
The International PRISM Task Force also has a significant European element, including leadership from the UK.

In the longer term (late 2020s onwards), the COMET collaboration will be also closely engaged with the next-generation PRISM experiment through the PRISM Task Force, which makes use of an FFAG muon storage ring to pursue detailed measurements of CLFV and LNV processes at J-PARC. 
This is a relatively long-term project which would be expected in the latter stages of the period relevant to the present strategy exercise.


\vspace{5mm}

%

\begin{sffamily} \begin{bfseries}
{\Large Summary}
\end{bfseries} \end{sffamily}

\bigskip

The COMET experiment is searching for the charged lepton flavour violating process of \mue conversion in a muonic atom at J-PARC in Japan. This search would be sensitive to BSM physics at energy scale of $\mathcal{O}$$(10^4)$ TeV. The COMET Phase-I experiment is under construction to seek a factor of 100 improvement over the current limit. By the mid-2020s, it is expected that the full COMET Phase-II will be deployed and running. It aims at a single event sensitivity of $2.6 \times 10^{-17}$ at $2 \times 10^{7}$ seconds of data-taking. Further improvements by one order of magnitude in sensitivity allowing to reach down to $\mathcal{O}(10^{-18})$ is likely. This improvement is based on refinements to the experimental design and operation. 
The COMET experiment has strong European participation, at about one in three of the membership of the collaboration. Their contributions and continued strong participation are crucial in making the COMET project a success.



\newpage
\pagenumbering{arabic}
%
%




\newpage
\begin{thebibliography}{}


\bibitem{muonCLFV} Muon CLFV submission ``Charged Lepton Flavour Violation using Intense Muon Beams at Future Facilities" to the European Strategy for 
Particle Physics 2020, and references therein.

\bibitem{TheMEG:2016wtm} 
  A.~M.~Baldini {\it et al.} [MEG Collaboration],
  Eur.\ Phys.\ J.\ C {\bf 76}, no. 8, 434 (2016).
  
\bibitem{Baldini:2018nnn} 
  A.~M.~Baldini {\it et al.} [MEG II Collaboration],
  Eur.\ Phys.\ J.\ C {\bf 78}, no. 5, 380 (2018).
  
\bibitem{Cavoto:2017kub} 
  G.~Cavoto, A.~Papa, F.~Renga, E.~Ripiccini and C.~Voena,
  Eur.\ Phys.\ J.\ C {\bf 78}, no. 1, 37 (2018).
  
\bibitem{Bellgardt:1987du} 
  U.~Bellgardt {\it et al.} [SINDRUM Collaboration],
  Nucl.\ Phys.\ B {\bf 299}, 1 (1988).
    
\bibitem{Blondel:2013ia} 
  A.~Blondel {\it et al.},
  arXiv:1301.6113 [physics.ins-det].

\bibitem{SINDRUM_MUEN} W. Bertl  {\it et al.} [SINDRUM-II Collaboration],  Eur.Phys.J. C47 337 (2006).
  
\bibitem{Nguyen:2015vkk} 
  T.~M.~Nguyen [DeeMe Collaboration],
  PoS FPCP {\bf 2015}, 060 (2015).
  
\bibitem{Bartoszek:2014mya} 
  L.~Bartoszek {\it et al.} [Mu2e Collaboration],
  arXiv:1501.05241 [physics.ins-det].

\bibitem{Mu2e-II-EOI} F. Abusalma {\it{et al.}} [Mu2e-II Experiment],
  arXiv:1802.02599 (2018).

\bibitem{comet-phaseI} R. Abramishvili {\it{et al.}} 
  [COMET Collaboration], \\
  http://comet.kek.jp/Documents\_files/PAC-TDR-2016/COMET-TDR-2016\_v2.pdf.

\bibitem{loi}  Y.~Kuno {\it et al.}  [COMET Collaboration], Letter of Intent (LOI) for ``An Experimental Search for the \mue Conversion at a Sensitivity of $10^{-16}$ with a Slow-Extracted Bunched Proton Beam",  December 15, (2006).

\bibitem{proposal}  D.~Bryman {\it et al.} [COMET Collaboration], Proposal for ``An Experimental Search for Lepton Flavor Violating \mue Conversion at a Sensitivity of $10^{-16}$ with a Slow-Extracted Bunched Proton Beam”.  November 30, (2007).

\bibitem{cdr}  Y.G.~Cui {\it et al.} [COMET Collaboration], Conceptual Design Report (CDR) for ``Experimental Search for Lepton Violating \mue Conversion at Sensitivity of $10^{-16}$ with a Slow-Extracted Bunched Proton Beam (COMET), J-PARC P21”. June 23, (2009).

\bibitem{Krikler:2016gdq}
  B.~Krikler,
  PhD Thesis, Imperial College London (2016).

\bibitem{technote164}
  Y.~Fujii {\it et al}, COMET Techical Note No 164 (2018), unpublished.
  
  
\bibitem{Crivellin:2017rmk} 
  A.~Crivellin, S.~Davidson, G.~M.~Pruna and A.~Signer,
  JHEP {\bf 1705}, 117 (2017).

\bibitem{Dohmen:1993mp} 
  C.~Dohmen {\it et al.} [SINDRUM II Collaboration],
  Phys.\ Lett.\ B {\bf 317}, 631 (1993).
  doi:10.1016/0370-2693(93)91383-X
 
\bibitem{Koike:2010xr} 
  M.~Koike, Y.~Kuno, J.~Sato and M.~Yamanaka,
  Phys.\ Rev.\ Lett.\  {\bf 105}, 121601 (2010).

\bibitem{Uesaka:2017yin} 
  Y.~Uesaka, Y.~Kuno, J.~Sato, T.~Sato and M.~Yamanaka,
  Phys.\ Rev.\ D {\bf 97}, no. 1, 015017 (2018).

\bibitem{Amhis:2016xyh} 
  Y.~Amhis {\it et al.} [HFLAV Collaboration],
  Eur.\ Phys.\ J.\ C {\bf 77}, no. 12, 895 (2017).
  
\bibitem{Lusiani:2018zvr} 
  A.~Lusiani,
  arXiv:1804.08436 [hep-ex].
  
\bibitem{MELC} R.M. Dzhilkibaev and V.M. Lobashev, Sov. J. Nucl. Phys. {\bf 49}, 384 (1989).
     
\bibitem{MUSIC} 
  S.~Cook {\it et al.},
  Phys.\ Rev.\ Accel.\ Beams {\bf 20}, no. 3, 030101 (2017).

\bibitem{Kitano:2002mt} 
  R.~Kitano, M.~Koike and Y.~Okada,
  Phys.\ Rev.\ D {\bf 66}, 096002 (2002),
  Erratum: [Phys.\ Rev.\ D {\bf 76}, 059902 (2007)].
  
\bibitem{Cirigliano:2009bz} 
  V.~Cirigliano, R.~Kitano, Y.~Okada and P.~Tuzon,
  Phys.\ Rev.\ D {\bf 80}, 013002 (2009).
    
\bibitem{Yeo:2017fej} 
  B.~Yeo, Y.~Kuno, M.~Lee and K.~Zuber,
  Phys.\ Rev.\ D {\bf 96}, no. 7, 075027 (2017).
      
\bibitem{Uesaka:2016vfy} 
  Y.~Uesaka, Y.~Kuno, J.~Sato, T.~Sato and M.~Yamanaka,
  Phys.\ Rev.\ D {\bf 93}, no. 7, 076006 (2016).
 
\bibitem{Cirigliano:2017azj} 
  V.~Cirigliano, S.~Davidson and Y.~Kuno,
  Phys.\ Lett.\ B {\bf 771}, 242 (2017).

\bibitem{Davidson:2017nrp} 
  S.~Davidson, Y.~Kuno and A.~Saporta,
  Eur.\ Phys.\ J.\ C {\bf 78}, no. 2, 109 (2018).
 
  

\end{thebibliography}
\end{document}